\documentstyle[aps,pra,epsbox]{revtex}
\begin{document}
\title{Effective Couplings of Dynamical Nambu-Goldstone Bosons with
Elementary Fermions}
\author{Takayuki Matsuki
\thanks{E-mail: matsuki@tokyo-kasei.ac.jp}}
\address{Tokyo Kasei University,
1-18-1 Kaga, Itabashi, Tokyo 173-8602, JAPAN}
\author{Masashi Shiotani
\thanks{E-mail: shiotani@tanashi.kek.jp}}
\address{Faculty of Science and Technology,
Kobe University, 1-1 Rokkohdai,\\ Nada, Kobe 657, JAPAN}
\maketitle
\medskip
\begin{abstract}
Assuming dynamical spontaneous breakdown of chiral symmetry for massless gauge
theory without scalar fields, we find a method how to construct an effective
action of the dynamical Nambu-Goldstone bosons and elemetary fermions by using
auxiliary fields. Here dynamical particles are asssumed to be composed of
elementary fermions. Various quantities including decay constants are
calculated from this effective action.\\
11.30.Rd, 11.15.Tk, 11.10.St
\end{abstract}
\section{Introduction}
\label{intro}
According to the paper by Nambu and Jona-Lasinio\cite{NJ} or more precisely the
Goldstone theorem,\cite{GS} if the Lagrangian without scalar fields has a
continuum global symmetry and if this symmetry is spontaneously broken, then
there must appear dynamically generated massless spin-0 partilcles, the
so-called Nambu-Goldstone (NG) bosons, the number of which is determined how
the symmetry is broken. Then it is necessary to have effective interations
among the NG bosons and elementary particles to study the effects of this
symmetry breakdown, either they appear as physical particles with tiny masses
or are absorbed by guage fields to make them massive.

Quite often these NG bosons acquire tiny masses through quantum corrections
and/or soft symmetry breaking to give actual physical effects. In these cases
we really need an effective Lagrangian to calculate those physical effects.
There are a couple of examples of effective Lagrangians known among the NG
bosons and elementary particles. For instance, there are the original
Nambu-Jona-Lasinio model\cite{NJ} of four-fermi interactions with spontaneous
chiral symmetry breakdown in which the NG boson is completely massless, and the
axion model\cite{WW} due to breakdown of the Peccei-Quinn symmetry\cite{PQ} in
which the axion (pseudo-NG boson) acquires a little mass through quantum
corrections, etc. The former case, which is the first paper to stimulate the
study of spontaneous symmetry breakdown, we just need to calculate the ordinary
Feynman rules for elementary fermions as well as would-be NG bosons which are
expressed as a chain of fermion loops. In the latter case we must assume
and/or guess effective interactions with some unknown coupling constants
although some constraints are enforced.

There is another interesting example in which people have calculated physical
quantities without an effective action. When one calculates the pion decay
constant, one uses the effective coupling between the pion (NG boson) and
quarks (or nucleons), which was, utilizing the Jackiw-Johnson\cite{JJ} sum
rule, originally derived by Pagels and Stokar\cite{PS} applying the
Ward-Takahashi (WT) identity to the axial vector vertex. This way of
determining the effective coupling, i.e., direct use of the WT identiy to the
vertex, includes ambiguities. The Kyoto group has used the Bethe-Salpeter (BS)
equation to uniquely determine this effective coupling and to calculate it
perturbatively.\cite{KT} They calculate the expectation value of
an axial vector current sandwitched between vacuum and pion state and relates
it with the BS amplitude to obtain the pion decay constant. In these approaches
there do not appear the dynamical NG bosons. Instead Pagels and Stoker use
only the WT identiy, and the Kyoto group uses a classical equation for the BS
amplitude. Hence it does not seem to be clear how actually the NG boson couples
to other elementary and/or dynamical fields as an operator form.

Here we propose one approach to construct an effective Lagrangian among the
NG bosons and elementary fields, i.e., the auxiliary field method, which was
once in fashion to construct various kinds of models out of four-fermi
interactions and also to study chiral symmetry breakdown.\cite{KSK,TA} With
this method one can determine uniquely the coupling of the NG boson with
fermions as an operator form. In this paper to show how powerful this approach
is, we start from the simplest example, $U(1)$ massless gauge theory without
scalar fields to derive the coupling among the dynamically generated  NG bosons
and fermions, to prove masslessness of the NG boson, and to calculate the decay
constant of this NG boson in Section \ref{U1}. Then in Section \ref{SU3} we
proceed into $SU(3)$ color gauge theory with iso-doublet fermions (up and down
quarks) without scalar fields to do the same thing in Section \ref{U1} and to
give the pion decay constant. The final Section \ref{disc} is devoted to the
summary and discussions of our method compared with others.
\section{$U(1)$ massless gauge theory}
\label{U1}
The Lagrangian for massless $U(1)$ gauge theory is given by
\begin{equation}
  {\cal L}_0=-\frac{1}{4}\left(F^{\mu\,\nu}\right)^2-\frac{1}{2\alpha}\left(
  \partial_\mu A^\mu\right)^2+\bar\psi\;i D{\kern -7pt /}\;\psi,\label{L0}
\end{equation}
where $D_\mu=\partial_\mu-igA_\mu$ and $\alpha$ is a gauge parameter. In this
paper we use Minkowski metric except for a few equations.
In order to introduce auxiliary fields as dynamical ones, we first
functionally integrate the partition function over gauge fields $A_\mu$.
The resultant four-fermi interaction terms are rewritten by using the Fierz
transformation as
\begin{eqnarray}
  S&=&\int\;d^4x\;\bar\psi(x)\;i \partial{\kern -6pt /}\;\psi(x)
  +\frac{1}{2}g^2\,\int\;d^4x\,d^4y\;\bar\psi(x)\gamma_\mu\psi(x)
  D^{\mu\,\nu}(x-y)\bar\psi(y)\gamma_\nu\psi(y)
  \nonumber \\
  &=&\int\;d^4x\;\bar\psi(x)\;i \partial{\kern -6pt /}\;\psi(x)
  -\frac{1}{2}\int\;d^4x\,d^4y\;\Big[\bar\psi(x)\psi(y)D(x-y)
  \bar\psi(y)\psi(x)
  \nonumber \\
  &+&\bar\psi(x)\;i\gamma_5\;\psi(y)D(x-y)\bar\psi(x)\;i\gamma_5
  \;\psi(y)\Big]+\ldots, \label{four}
\end{eqnarray}
where $\ldots$ expresses other modes and
\begin{mathletters}
\label{D}
\begin{eqnarray}
  D_{\mu\,\nu}(x-y)&=&\int\frac{d^4p}{\left(2\pi\right)^4}\frac{e^{-ip(x-y)}}
  {p^4}\left[p^2g_{\mu\,\nu}-\left(1-\alpha\right)p_\mu p_\nu\right],\\
  D(x-y)&=&\frac{g^2}{4}\,g^{\mu\,\nu}D_{\mu\,\nu}(x-y)=
  \frac{3+\alpha}{4}\,g^2\int\frac{d^4p}{\left(2\pi\right)^4}
  \frac{e^{-ip(x-y)}}{p^2}=-\frac{\lambda}{\left(x-y\right)^2},\\
  \lambda&=&\frac{(3+\alpha)g^2}{16\pi^2}. \label{lambda}
\end{eqnarray}
\end{mathletters}

Then by introducing the gaussian term which is quadratic in auxiliary fields
and cancels four-fermi terms in Eq.~(\ref{four}) as
\begin{eqnarray}
  S_{\rm AF}&=&\frac{1}{2}\int\;d^4x\,d^4y\;\left[\phi(x,y)-\bar\psi(x)\psi(y)
  D(x-y)\right]D^{-1}(x-y)
  \nonumber \\
  &\quad&\times\left[\phi(y,x)-D(y-x)\bar\psi(y)\psi(x)\right]
  \nonumber \\
  &+&\frac{1}{2}\int\;d^4x\,d^4y\;\left[\pi(x,y)-\bar\psi(x)\;i\gamma_5\;
  \psi(y)D(x-y)\right]D^{-1}(x-y)
  \nonumber \\
  &\quad&\times\left[\pi(y,x)-D(y-x)\bar\psi(y)\;i\gamma_5\;\psi(x)\right]
  +\ldots,
\end{eqnarray}
we obtain\cite{TA},
\begin{eqnarray}
  S+S_{\rm AF}&=&\int\;d^4x\;\bar\psi(x)\;i \partial{\kern -6pt /}\;\psi(x)
  -\int\;d^4x\,d^4y\;\psi(x)\Big[\phi(x,y)+i\gamma_5\pi(x,y)\Big]\psi(y)
  \nonumber \\
  &+&\frac{1}{2}\int\;d^4x\,d^4y\;\Big[\phi(x,y)D^{-1}(x-y)\phi(y,x)+
  \pi(x,y)D^{-1}(x-y)\pi(y,x)\Big]+\ldots, \label{Yukawa}
\end{eqnarray}
where $D^{-1}(x-y)$ is nothing but $1/D(x-y)$ and assumed are
$\phi(x,y)=\phi(y,x)$ and $\pi(x,y)=\pi(y,x)$. Since main purpose of this paper
is to present our idea as simple as possible, we have neglected other modes.

Next we assume that bilocal fields can be decomposed into local fields with
continuum indices as follows, which is essential to the following whole
arguments in this paper.
\begin{mathletters}
\label{phi-pi1}
\begin{eqnarray}
  \phi(x,y)&=&\phi_0(x-y)+\frac{\sigma_0(x-y)}{f}\;\sigma\left(\frac{x+y}{2}
  \right),\\
  \pi(x,y)&=&\frac{\pi_0(x-y)}{f}\;\varphi\left(\frac{x+y}{2}\right),\label{pi}
\end{eqnarray}
\end{mathletters}
where fields with arguments $x-y$ are considered to be classical fields
and those with $(x+y)/2$ to be quantum ones like in the operator product
expansion. Also assumed is that $\phi_0(x-y)$ is a vacuum expectation value
(VEV) of $\phi(x,y)$. The VEV should depend only on $x-y$ because of
translational invariance. The coordinate $(x+y)/2$ is the center of mass
system. Here a constant parameter $f$ has massive dimension one,
$\phi_0(x-y)$, $\sigma_0(x-y)$, and $\pi_0(x-y)$ dimension five, and
$\sigma((x+y)/2)$ and $\varphi((x+y)/2)$ dimension one. The difference
between quantum and classical fields can be seen by Fourier-transforming fields
with arguments $x-y$ into momentum space, in which case, e.g., $\pi(x,y)$ can
be interpreted as local fields with continuum label of internal momentum.
That is,
\begin{eqnarray*}
  \pi(x,y)=\frac{\pi_0(x-y)}{f}\;\varphi\left(\frac{x+y}{2}\right) 
  =\frac{1}{f}\int \frac{d^4q}{(2\pi)^4}\,e^{-iqr}\,\tilde\pi_0(q)\,
  \varphi\left(X\right),
\end{eqnarray*}
with $r=x-y$ and $X=(x+y)/2$, which shows $\pi(x,y)$ is equivalent to
$\tilde\pi_0(q)\,\varphi\left(X\right)$ with a continuum label $q$.

The legitimacy of this decomposition Eqs.~(\ref{phi-pi1}) may be supported by
the following observation. The composite field in question is tightly bound due
to the strong coupling which causes chiral symmetry breakdown and hence it can
be decomposed into two wave functions as the first approximation, one for
internal motion and another for total motion. An internal motion can be
described in $x-y$, difference of coordinates of two particles, and the whole
motion in $(x+y)/2$, the center of mass system coordinate of two particles.
Anyway this is the simplest decomposition of $\phi(x,y)$ and $\pi(x,y)$. This
could also be justified by checking whether mass for the would-be NG boson,
$\varphi\left((x+y)/2\right)$ becomes massless or not which we will see soon.

From symmetrical point of view (rotational invariance), it holds
\begin{equation}
  \sigma_0(x-y)=\pi_0(x-y).
\end{equation}
Since we are interested in describing a coupling of the NG boson, i.e.,
$\varphi((x+y)/2)$, with fermions and in obtaining the equations for
$\phi_0(x-y)$ and $\pi_0(x-y)$, functionally integrating over fermions and
keeping only $\phi_0(x-y)$, $\pi_0(x-y)$, and $\varphi((x+y)/2)$, we obtain
\begin{eqnarray}
  S_{\rm eff}&=&\frac{1}{2}\int\,d^4X\, d^4r\;\left[\phi_0(r)D^{-1}(r)
  \phi_0(r)+\frac{1}{f^2}\left(\varphi(X)\right)^2\pi_0(r)D^{-1}(r)
  \pi_0(r)\right]
  \nonumber \\
  &-&i\,{\rm Tr}\,\ln\,\Big[i\partial{\kern -6pt /}_r-\phi_0(r)-
  \frac{i}{f}\pi_0(r)\gamma_5\;\varphi(X)\Big],
  \nonumber \\
  &=&\frac{1}{2\lambda}\int\,d^4X\,\frac{d^4q}{\left(2\pi\right)^4}\;\Biggl\{
  \Sigma(q)\partial_q^2\Sigma(q)+\frac{1}{f^2}\left(
  \varphi(X)\right)^2\tilde\pi_0(q)\partial_q^2\tilde\pi_0(q)
  \nonumber \\
  &-&4i\lambda\,\ln\,\left[q^2-\Sigma^2(q)-\frac{1}{f^2}\,
  \tilde\pi_0^2(q)\left(\varphi(X)\right)^2\right]\Biggr\}, \label{Seff}
\end{eqnarray}
where $\lambda$ is given by Eq.~(\ref{lambda}), Tr is to take trace over
coordinates ($x$ and $y$ or $r=x-y$ and $X=(x+y)/2$ or $X$ and $q$ where $q$
is a conjugate momentum to $r$) and gamma-matrix indices, $\partial_q^2$ is a
dalambertian in terms of $q_\mu$ in Minkowski space and
\begin{equation}
  \Sigma(q)=\int \frac{d^4r}{\left(2\pi\right)^4}\phi_0(r)\;e^{-iqr},
  \qquad
  \tilde\pi_0(q)=\int \frac{d^4r}{\left(2\pi\right)^4}\pi_0(r)\;e^{-iqr}.
\end{equation}
In calculating Tr~$\ln$, we have negelected the term $\partial{\kern -6pt /}_X$
inside since we are not interested in the derivative terms in $X$.

Now setting $\varphi(X)=0$ in Eq.~(\ref{Seff}) since the VEV of this field is
zero by definition ($\varphi(X)$ is the NG boson) and varying $S_{\rm eff}$ in
terms of $\Sigma(q)$, we can obtain an equation for $\Sigma(q)$. This is
nothing but the gap equation, i.e., this gives the mass, $\Sigma(q)$, for a
fermion.
\begin{equation}
  \partial_q^2\Sigma(q)+\frac{4i\lambda\Sigma(q)}{q^2-\Sigma^2(q)}=0,
  \label{gap1}
\end{equation}
or
\begin{equation}
  \Sigma(q)=4i\lambda\,\int\,\frac{d^4p}{(2\pi)^2}\frac{1}{(p-q)^2}
  \frac{\Sigma(p)}{p^2-\Sigma^2(p)},
  \label{gap2}
\end{equation}
which becomes a familiar form when converted into one in Euclidean space and
in the Landau gauge $\alpha=0$ or $\lambda=3g^2/(16\pi^2)$. Here we have used
\begin{equation}
  \left(\partial_q^2\right)^{-1}=-\int\,\frac{d^4p}{(2\pi)^2}\frac{1}{(p-q)^2}.
\end{equation}

Next extracting only the bilinear terms in $\varphi(X)$ from Eq.~(\ref{Seff}),
we obtain the mass term for $\varphi(X)$ as
\begin{equation}
  \frac{1}{2\lambda f^2}\int\;d^4X\,\frac{d^4q}{\left(2\pi\right)^4}
  \left(\varphi(X)\right)^2\tilde\pi_0(q)\left[\partial_q^2+
  \frac{4i\lambda}{q^2-\Sigma^2(q)}\right]\tilde\pi_0(q),
\end{equation}
which is identically zero when one identifies $\tilde\pi_0(q)$ with $\Sigma(q)$
and uses Eq.~(\ref{gap1}). In other words, when one requires masslessness of the
NG boson $\varphi(X)$, then $\tilde\pi_0(q)=\Sigma(q)$ is automatically derived
up to a constant $f$. This {\it proves} masslessness of the NG boson,
$\varphi(X)$. Back to Eq.~(\ref{Yukawa}) this result means that the Yukawa
coupling of the NG boson $\varphi(X)$ with fermions is given by
\begin{eqnarray}
  &&-\int\;d^4x\,d^4y\;\bar\psi(x)\;i\gamma_5\;\pi\left(x,y\right)\;\psi(y)=
  \nonumber \\
  &&\qquad-\int\;\frac{d^4p}{\left(2\pi\right)^4}\frac{d^4q}
  {\left(2\pi\right)^4}\bar{\tilde\psi}(p+q)\frac{i}{f}\gamma_5\Sigma
  \left(p+\frac{q}{2}\right)\tilde\varphi(q)\tilde\psi(p),
\end{eqnarray}
in momentum space. Here fields with tilde are Fourier-transformed ones.
The Feynman rule for the vertex of $\bar\psi\varphi\psi$, i.e.,
fermion-anti-fermion-NG boson vertex, is finally given by
\begin{equation}
  \frac{1}{f}\Sigma\left(p+\frac{q}{2}\right)\gamma_5. \label{vertex}
\end{equation}
This correponds to the Feynman diagram depicted in Fig.1 with a vertex factor
given by Eq.(\ref{vertex}).

\begin{center}
%
%
\begin{figure}
\center\psbox[hscale=0.5,vscale=0.5]{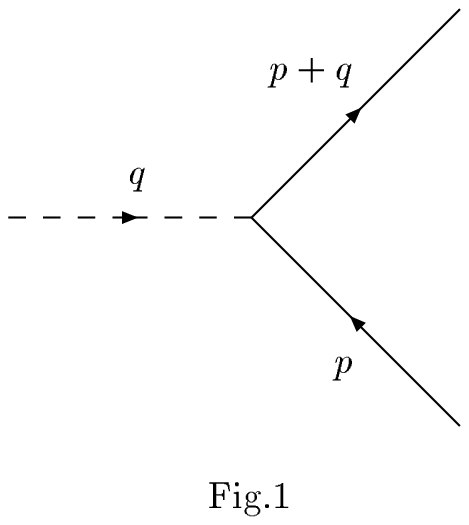}
\caption{Bare vertex of $\bar\psi\varphi\psi$.}
\label{Fig1}
\end{figure}
\end{center}

This result coincides with that derived by using the BS equation\cite{KT} and
slightly differs from that originally derived by Pagels and Stokar\cite{PS}
using the WT identity. Eqs.~(\ref{phi-pi1}) are rewritten as
\begin{mathletters}
\label{phi-pi2}
\begin{eqnarray}
  \phi(x,y)&=&\frac{\phi_0(x-y)}{f}\left[f+\sigma\left(\frac{x+y}{2}
  \right)\right], \label{phi} \\
  \pi(x,y)&=&\frac{\phi_0(x-y)}{f}\;\varphi\left(\frac{x+y}{2}\right).
\end{eqnarray}
\end{mathletters}
These equations mean that the fields $\phi(x,y)$ and $\pi(x,y)$ should be
rescaled by a factor $\phi_0(x-y)/f$ to obtain ordinary dynamical fields and
a scalar field $\sigma'(X)\;(=f\;\phi(x,y)/\phi_0(r))$ has a vacuum
expectation value $f$ to become a massive scalar field $\sigma(X)$.

Now we calculate the decay constant from this vertex. The decay constant in the
$U(1)$ case is given by $2f$ and is defined by a matrix element of the axial
vector current as
\begin{equation}
  \left<0\left|\bar\psi(0)\;\gamma_\mu\gamma_5\psi(0)\right|\varphi(q)\right>=
  -2i\,f\,q_\mu. \label{Dconst}
\end{equation}
In the $q_\mu\rightarrow 0$ limit, our vertex, the Beth-Salpeter vertex, and
the Pagels-Stokar vertex, all give the same expression for the decay constant
$f$. The factor 2 of Eqs.~(\ref{Dconst}) is explained in the final Section
\ref{disc}, wich is peculiar to $U(1)$ theory. The Feynman diagram which
corresponds to the left hand side (lhs) of Eq.~(\ref{Dconst}) is depicted
in Fig.2 which gives, keeping only a linear term in momentum $q_\mu$ and with
the help of Eq.~(\ref{vertex}), the
final answer as,
\begin{equation}
  f^2=\frac{1}{2\left(2\pi\right)^2}\,\int\,xdx\,
  \frac{\Sigma(x)\Bigl[\Sigma(x)-x\Sigma'(x)/2\Bigr]}
  {\Bigl[x+\Sigma^2(x)\Bigr]^2},
\end{equation}
where momentum space is converted into Eucledean space, i.e., $x=-p^2$. Be
careful when one converts the results of Fig.2 to the r.h.s. of
Eq.~(\ref{Dconst}) bacause $q_\mu$ is an outgoing momentum.

\begin{center}
%
%
\begin{figure}
\center\psbox[hscale=0.5,vscale=0.5]{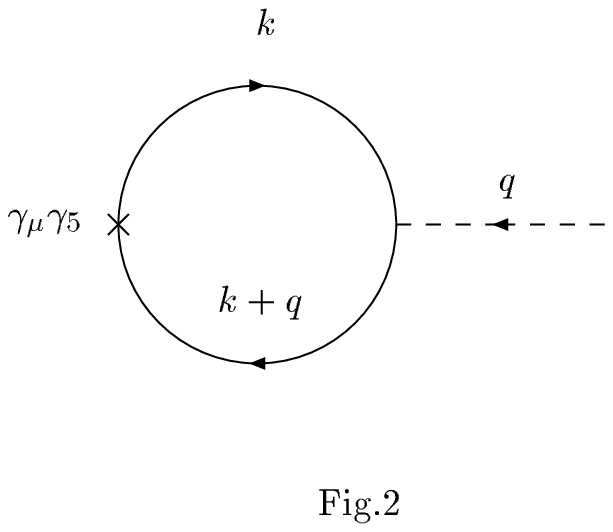}
\caption{Fermion one loop correction to the axial vector current.}
\label{Fig2}
\end{figure}
\end{center}

\section{$SU(3)$ color gauge theory with iso-doublet}
\label{SU3}
Now that we have described our idea in the former section, we would like to
proceed into a more realistic case, $SU(3)$ color gauge theory with massless
iso-doublet fermions, $u$ and $d$ quarks.

The difference between $U(1)$ and $SU(3)$ cases is first the color factor, and
secondly the flavor number. Although the arguments to obtain the pion decay
constant $f_\pi$ in $SU(3)$ are almost parallel to the $U(1)$ case, there are
a couple of numerical factor differences. Hence only the equations different
from $U(1)$ case are described below. First we have the following coupling
beween the dynamical scalar and fermions,
\begin{equation}
  -\int\;d^4x\,d^4y\;\psi(x)\Big[\phi(x,y)+i\gamma_5\pi^a(x,y)\tau^a
  \Big]\psi(y),
\end{equation}
where $\psi(x)=\left(u(x),d(x)\right)^{\rm T}$ and $\tau^a$ are the $2\times 2$
Pauli matrices. Likewise we can easily derive masslessness of the NG boson,
$\varphi^a(X)$, and the auxiliary fields given by Eqs.~(\ref{phi-pi1}) should
be replaced with
\begin{mathletters}
\label{phi-pi3}
\begin{eqnarray}
  \phi(x,y)&=&\frac{\phi_0(x-y)}{f_\pi}\left[f_\pi+\sigma\left(\frac{x+y}{2}
  \right)\right], \label{phi2} \\
  \pi^a(x,y)&=&\frac{\phi_0(x-y)}{f_\pi}\;\varphi^a\left(\frac{x+y}{2}\right),
\end{eqnarray}
\end{mathletters}
with
\begin{equation}
  \int \frac{d^4r}{\left(2\pi\right)^4}\phi_0(r)\;e^{-iqr}=\Sigma(q).
\end{equation}
Instead of Eq.~(\ref{Dconst}) for the $U(1)$ axial vector current, the
$SU(3)$ axial vector iso-doublet current satisfies
\begin{equation}
  \left<0\left|\bar\psi(0)\;\gamma_\mu\gamma_5\frac{\tau^a}{2}\psi(0)
  \right|\varphi^a(q)\right>=-i\,f_\pi\,q_\mu. \label{Dconst2}
\end{equation}
Note that the coefficient of the rhs of Eq.~(\ref{Dconst2}) is 1 instead of 2.
The Feynman rule for the vertex $\bar\psi\varphi^a\psi$ in momentum space is
given by
\begin{equation}
  \frac{1}{f_\pi}\Sigma\left(p+\frac{q}{2}\right)\gamma_5\,\tau^a,
\end{equation}
and we only give the final expression for $f_\pi$ as
\begin{equation}
  f^2_\pi=\frac{N_c}{\left(2\pi\right)^2}\,\int\,xdx\,
  \frac{\Sigma(x)\Bigl[\Sigma(x)-x\Sigma'(x)/2\Bigr]}
  {\Bigl[x+\Sigma^2(x)\Bigr]^2},
\end{equation}
where $N_c=3$ is a number of colors. Only be careful that the Fierz
transformation of internal degrees of freedom, i.e., $SU(2)$ flavor as well as
$SU(3)$ color must be taken into account, in which case use has been made of
\begin{mathletters}
\begin{eqnarray}
  \delta_{\alpha\,\beta}\,\delta_{\delta\,\gamma}&=&\frac{1}{N}
  \delta_{\alpha\,\gamma}\,\delta_{\delta\,\beta}+2
  {\left(T^a\right)_{\alpha\,\gamma}}\,
  {\left(T^a\right)_{\delta\,\beta}}, \\
  {\left(T^a\right)_{\alpha\,\beta}}\,
  {\left(T^a\right)_{\delta\,\gamma}}&=&\frac{1}{2}
  \left(1-\frac{1}{N^2}\right)-\frac{4}{N}\,
  {\left(T^a\right)_{\alpha\,\gamma}}\,
  {\left(T^a\right)_{\delta\,\beta}}, \\
  {\rm Tr}\,T^a&=&0,\quad {\rm Tr}\,\left(T^a\,T^b\right)=\frac{1}{2}
  \delta_{a\,b},
\end{eqnarray}
\end{mathletters}
where $T^a$ are generators and $N$ is a number of dimensions of the group.
\section{Summary and Discussion}
\label{disc}
In this paper we have shown how useful and easy our approach is and derived
that by introducing the auxiliary fields, composite fields have a coupling with
fermions and can be decomposed into local fields multiplied by a factor which
is given by the solution to the gap equation, that the NG bosons are acually
massless, and that the decay constant $f$ is a vacuum expectation value of a
composite scalar field and calculated by a one-loop fermion diagram given by
Fig.2 after taking a limit of $q_\mu\rightarrow 0$. These have been done both
for $U(1)$ and $SU(3)$ massless gauge theories without scalar fields.

Now in this section, we show the relation in the simplest $U(1)$ case between
our method without the WT identity and others using the WT identity. If the
chiral symmetry is not broken, the axial vector Ward-Takahashi identity is
given by
\begin{equation}
  q^\mu\Gamma^5_\mu(p+q,p)=S_F^{-1}(p+q)\;\gamma_5+
  \gamma_5\;S_F^{-1}(p), \label{WT1}
\end{equation}
where $S_F(p)$ is a fermion full propagator given by
\begin{equation}
  S_F(p)=\frac{i}{p{\kern -4.5pt /}-\Sigma(p)}.
\end{equation}

The usual way to obtain Eq.~(\ref{WT1}) is to calculate the divergence of the
T product,
\begin{mathletters}
\begin{eqnarray}
  \int\,d^4x\,e^{-iqx-ipy}&\partial^\mu&\left<0\right|{\rm T}\left(J^5_\mu(x)
  \psi(y)\bar\psi(0)\right)\left|0\right>, \label{Tproduct} \\
  J^5_\mu(x)&=&\bar\psi(x)\gamma_\mu\gamma_5\psi(x)=
  -\frac{\delta {\cal L}}{\delta\partial^\mu\psi}\,\delta\psi.\label{Noether1}
\end{eqnarray}
\end{mathletters}
Here $J^5_\mu(x)$ is the Noether current for the chiral symmetry.
Eq.~(\ref{Tproduct}) can be estimated by using the following anti-commutation
relation, transformation laws, and conservation equation for the chiral
symmetry,
\begin{mathletters}
\label{conserve1}
\begin{eqnarray}
  \left\{\psi^\dagger(\vec x,x^0),\,\psi(\vec y,y^0)\right\}\Big|_{\,x^0=y^0}
  &=&i\,\delta^3(\vec x-\vec y), \\
  \left[Q^5,\psi(x)\right]=i\gamma_5\psi(x),&\quad&
  \left[Q^5,\bar\psi(x)\right]=i\bar\psi(x)\gamma_5,\quad
  Q^5=\int d^3x\,J^5_0(x),\label{Qpsi} \\
  \partial^\mu\,J^5_\mu(x)&=&0,
\end{eqnarray}
\end{mathletters}
where $Q^5$ is an operator which generates chiral transformation. Negative sign
of Eq.~(\ref{Noether1}) comes from Eqs.~(\ref{Qpsi}). Now let us
graphically calculate $\Gamma^5_\mu(p+q,p)$ when the chiral symmetry is
spontaneously broken. In this case, since only the fermions as well as gauge
fields are elementary, a massless pole contribution to $\Gamma^5_\mu(p+q,p)$
can be depicted in Fig.3 and the whole $\Gamma^5_\mu(p+q,p)$ with a tree vertex
is given by,
\begin{mathletters}
\begin{eqnarray}
  \,\Gamma^5_\mu(p+q,p)&=&i\,\gamma_\mu\gamma_5-2i\,f\,
  G(p+q,p)\frac{q_\mu}{q^2}+{\rm regular~part}, \label{WT2} \\
  G(p+q,p)&=&\frac{1}{f}\Sigma\left(p+\frac{q}{2}\right)\,\gamma_5,\label{Gpq}
\end{eqnarray}
\end{mathletters}
where "regular part" means the remaining terms which are not singular at
$q^2=0$.

\begin{center}
%
%
\begin{figure}
\center\psbox[hscale=0.5,vscale=0.5]{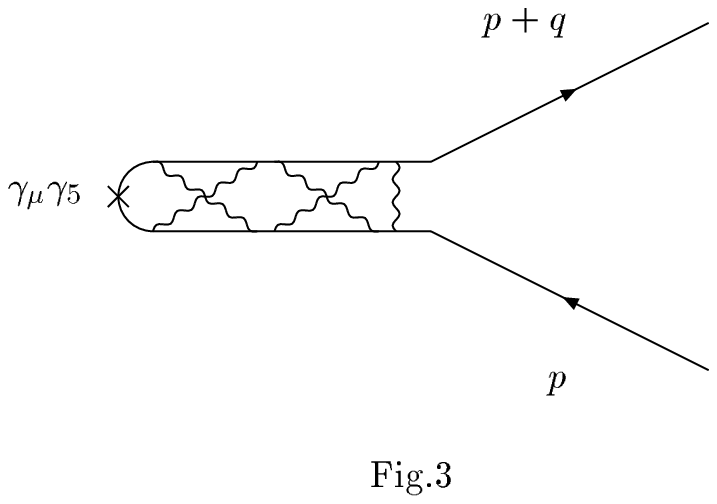}
\caption{Massless pole correction to the axial-vector vertex.}
\label{Fig3}
\end{figure}
\end{center}

These are the arguments of References \cite{JJ,PS,KT}. Pagels and Stokar
\cite{PS} have suggested the form of $G(p+q,p)$ so that the WT identiy
Eqs.~(\ref{WT1}, \ref{WT2}) is satisfied and gives a form slightly different
from Eq.~(\ref{Gpq}). The Kyoto group \cite{KT} uses Eq.~(\ref{Gpq}) as a
normalization of the BS amplitude. All these arguments have meaning only in the
limit of $q_\mu\rightarrow 0$. We would like to put stress upon again the
reason why the second term in the rhs of Eq.~(\ref{WT2}) is added to
$\Gamma^5_\mu(p+q,p)$. This is because people so far have treated the NG boson
as a composite particle of fermions, not an elementary field, and hence the
Feynman diagram for the vertex shown in Fig.3 consisting only of fermions and
gauge fields should be included.

On the other hand, in our case we treat the NG boson as an elementary field
although dynamical so that we can directly derive the coupling between fermions
and the NG boson. Let us see how this affects the above arguments on
derivation of the WT identity. Since the NG boson as well as a scalar field
are introduced in the effective Lagrangian, the Noether current must be
modified so that it includes the contributions from the scalars, which is
given by
\begin{equation}
  \tilde J_\mu^5(x)=J_\mu^5(x)-\frac{\delta {\cal L}}
  {\delta\partial^\mu\varphi}\,\delta\varphi-\frac{\delta {\cal L}}
  {\delta\partial^\mu\sigma'}\,\delta\sigma',
  \label{Noether2}
\end{equation}
where $J_\mu^5(x)$ is given by Eq.~(\ref{Noether1}). In our case the correct
Noether current is given by Eq.~(\ref{Noether2}) and hence replacing
$J_\mu^5(x)$ with $\tilde J_\mu^5(x)$ in Eqs.~(\ref{conserve1}), the following
equations hold,
\begin{mathletters}
\label{conserve2}
\begin{eqnarray}
  \left[\tilde Q^5,\psi(x)\right]=i\gamma_5\psi(x),&\quad&
  \left[\tilde Q^5,\bar\psi(x)\right]=i\bar\psi(x)\gamma_5,\\
  \left[\tilde Q^5,\sigma'(x)\right]=2\varphi(x),&\quad&
  \left[\tilde Q^5,\varphi(x)\right]=-2\sigma'(x),\label{charge} \\
  \tilde Q^5=\int d^3x\,\tilde J^5_0(x),&\quad&
  \partial^\mu\,\tilde J^5_\mu(x)=0.
\end{eqnarray}
\end{mathletters}
Using these equations, one can easily show that the same WT identity as
Eq.~(\ref{WT1}) holds since Eqs.~(\ref{charge}) do not affect derivation of this equation.

The second and third terms of the rhs of Eq.~(\ref{Noether2}) are expected to
be obtained from the kinetic terms of the scalars, which are, however, not
present in the effective Lagrangian obtained after a couple of manipulations.
Therefore either that kinetic terms must be generated by quantum corrections or
the correction terms to the Noether current should directly be calculated. What
we have done in Sect. \ref{U1} is a calculation of the correction term directly
concerned with the chiral symmetry breakdown. There we have calculated Fig.2 and obtained the term proportional to $\partial_\mu\varphi$ that we need now.
That is, we have required that the diagram should satify
\begin{equation}
  -\frac{\delta {\cal L}}{\delta\partial^\mu\varphi}\,\delta\varphi=
  2f\partial_\mu\varphi(x)+\ldots,
\end{equation}
which determines $f$ in terms of $\Sigma(x)$ after symmetry breaks down. Namely
the Feynman diagram corresponding to Fig.3 is now given by Fig.4, which gives
massless particle (NG boson) contribution to the vertex. This is a
multiplication of two diagrams, Fig.2 and Fig.1 and in between them there is a
massless NG boson propagator which gives a $1/q^2$ pole term.

\begin{center}
%
%
\begin{figure}
\center\psbox[hscale=0.5,vscale=0.5]{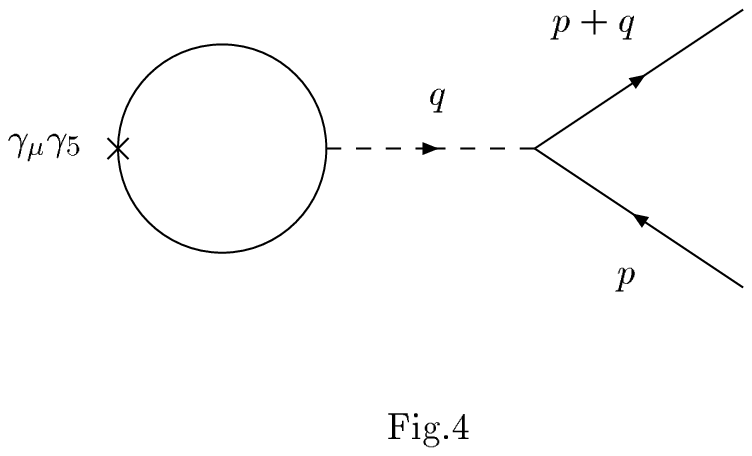}
\caption{Dynamical Nambu-Goldstone boson contribution to the axial vector
vertex.}
\label{Fig4}
\end{figure}
\end{center}

Fig.2 gives a factor $-2f\,q_\mu$, Fig.1 does $\Sigma(p+q/2)\,\gamma_5/f$, and
the $\varphi$ propagator does $i/q^2$. After multiplying all these factors, we
obtain $-2i\Sigma(p+q/2)\,\gamma_5\,q_\mu/q^2$, which is exactly equal to the
second term of the rhs of the WT identity Eq.~(\ref{WT2}).

Finally we would like to describe another simple method to calculate the decay
constant or $f$. In order to estimate Eq.~(\ref{Dconst}) we lift global chiral
symmetry to local one. That is, we introduce gauge fields associated with
local chiral symmetry, calculate some diagram to obtain $f$, and then turn off
local symmetry to have the original global chiral symmetry. The advantage to
have local symmetry is to have a gauge coupling term with the Noether current,
$\bar\psi(x)\;\gamma^\mu\gamma_5\psi(x)$, as
\begin{equation}
  A^5_\mu(x)\,\bar\psi(x)\;\gamma^\mu\gamma_5\psi(x), \label{A5}
\end{equation}
which helps calculate $f$. We here neglect the chiral anomaly as well as its
effects due to introduction of local chiral symmtery or $A^5_\mu(x)$. This does
not affect the final results. The fermion-anti-fermion-NG boson vertex is given
by
\begin{equation}
  \tilde\Gamma^5_\mu(p+q,p)=\int d^4x\,d^4y \,e^{-iqx-ipy}\left<0\right|
  {\rm T}\left(A^5_\mu(x) \psi(y) \bar\psi(0)\right)\left|0\right>_{\rm amp},
\end{equation}
which corresponds to a bare vertex $i\gamma_\mu\gamma_5$. Here the subscript
"amp" means vertex functions are amputated. Taking an analogy to the ordinary
elementary Higgs field, what we need is the following two-point function,
\begin{equation}
  \tilde\Gamma_\mu(q)=\int d^4x \,e^{-iqx}\left<0\right|
  {\rm T}\left(A^5_\mu(0) \varphi(x)\right)\left|0\right>_{\rm amp},
  \label{gaugePhi}
\end{equation}
which is nothing but $2f\,q_\mu$ in the limit of $q_\mu\rightarrow 0$ which
will be shown below. With the help of Eq.~(\ref{A5}) Eq.~(\ref{gaugePhi}) is
given by a fermion one-loop diagram depicted in Fig.5 and equating this to
$2f\,q_\mu$ gives the expression for $f$ in terms of $\Sigma(x)$. This is the
simplest way to obtain the Pagels-Stokar formula compared with other methods.

\begin{center}
%
%
\begin{figure}
\center\psbox[hscale=0.5,vscale=0.5]{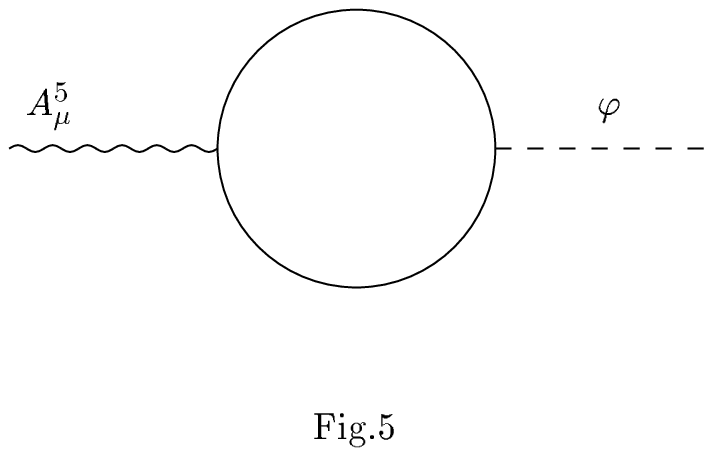}
\caption{Fermion loop correction to the $A^5_\mu-\varphi$ two point function.}
\label{Fig5}
\end{figure}
\end{center}

To see that the above two-point function Eq.~(\ref{gaugePhi}) gives
$2f\,q_\mu$, we expand the ordinary "gauge-invariant" Higgs kinetic term,
\begin{eqnarray}
  &\frac{1}{2}&\left|D_\mu\left(\sigma'(x)+i\varphi(x)\right)
  \right|^2 \nonumber \\
  &=\frac{1}{2}&\left[\left(\partial^\mu-2iA^{5\,\mu}\right)\left(
  \sigma'(x)-i\varphi(x)\right)\left(\partial_\mu+2iA^5_\mu\right)
  \left(\sigma'(x)+i\varphi(x)\right)\right]=2f\,A^5_\mu\partial^\mu
  \varphi+\ldots, \label{kinetic}
\end{eqnarray}
where the charge of the scalar term $\sigma'(x)+i\varphi(x)$ is -2 as shown by
Eq.~(\ref{charge}) and use has been made of $<0|\sigma'(x)|0>=f$ as given by
Eq.~(\ref{phi}). Notice that a relative sign between
$A_\mu^5\bar\psi\gamma_\mu\gamma_5\psi$ and $2fA_\mu^5\partial^\mu\varphi$
in Eq.~(\ref{kinetic}) is plus which is  the same as in Eq.~(\ref{Noether2}).

There are some models in which there are still gauge fields remained after
functionally integrating a couple of strongly coupled gauge fields. In this case
there remains local symmetry and we have terms similar to Eq.~(\ref{A5}) which
give, as we have seen, masses to remained gauge fields, which will be a next
problem to be solved.

\acknowledgments
The authors would like to thank the theory group of KEK at Tanashi where part
of this work has been done. One of the authors (TM) also acknowledges
Prof. R. Haymaker for his hospitality at Louisiana State University where the
final work has been done.

\end{document}